\input ./style/arxiv-general.cfg
\documentclass[aoas,MSNbibl,nameyear,seceqn,dvips]{arximspdf}
\makeatletter
   \@ifpackageloaded{graphicx}{}{\usepackage{graphicx}}
\makeatother
\usepackage{mathrsfs,mathbh}
\usepackage{dcolumn}

\doi{10.1214/15-AOAS852}
\volume{9}
\issue{3}
\pubyear{2015}
\firstpage{1621}
\lastpage{1642}
\docsubty{FLA}

\makeatletter
\newcolumntype{d}[1]{D{.}{.}{#1}}
\def\mathbbm{\mathbh}
\def\cal{\mathcal}

\newproclaim{df}[lm]{Definition}

\newproclaim{remark}[lm]{Remark}

\def\diag{\operatorname{diag}}
\def\wh{\widehat}
\newcommand{\CJ}{{\cal J}}
\newcommand{\CO}{{\cal O}}
\newcommand{\CT}{{\cal T}}
\newcommand{\SB}{\mathscr{B}}
\newcommand{\MX}{{\mathbb X}}

\newcommand{\E}{{\mathrm E}}

\def\curr{\mathrm{curr}}

\def\diag{\operatorname{diag}}
\def\wh{\widehat}
\def\log{\operatorname{log}}
\def\Normal{\operatorname{Normal}}
\def\wh{\widehat}

\makeatother

\begin{document}
\begin{frontmatter}

\title{Joint modeling of longitudinal drug using pattern and time to
first relapse in cocaine dependence treatment data}
\runtitle{Joint modeling in cocaine dependence treatment data}

\begin{aug}
\author[A]{\fnms{Jun} \snm{Ye}\corref{}\thanksref{T1}\ead[label=e1]{jye1@uakron.edu}},
\author[B]{\fnms{Yehua} \snm{Li}\thanksref{T2}\ead[label=e2]{yehuali@iastate.edu}}
\and
\author[C]{\fnms{Yongtao} \snm{Guan}\thanksref{T3}\ead[label=e3]{yguan@bus.miami.edu}}
\runauthor{J. Ye, Y. Li and Y. Guan}
\thankstext{T1}{Supported in part by 2014 University of Akron BCAS
Faculty Scholarship Award.}
\thankstext{T2}{Supported in part by NSF Grants DMS-11-05634 and DMS-13-17118.}
\thankstext{T3}{Supported in part by NIH Grant 5R01DA029081-04 and NSF
Grant DMS-08-45368.}
\affiliation{University of Akron, Iowa State University and University
of Miami}
\address[A]{J. Ye\\
Department of Statistics\\
University of Akron\\
Akron, Ohio 44325\\
USA\\
\printead{e1}}

\address[B]{Y. Li\\
Department of Statistics\\
\quad and Statistical Laboratory\\
Iowa State University\\
Ames, Iowa 50011\\
USA\\
\printead{e2}}

\address[C]{Y. Guan\\
Department of Management Science\\
School of Business Administration\\
University of Miami \\
Coral Gables, Florida 33124\\
USA\\
\printead{e3}}
\end{aug}

%
\received{\smonth{9} \syear{2014}}
%
\revised{\smonth{5} \syear{2015}}

%
\begin{abstract}
An important endpoint variable in a cocaine rehabilitation study is the
time to first relapse of a patient after the treatment. We propose a
joint modeling approach based on functional data analysis to study the
relationship between the baseline longitudinal cocaine-use pattern and
the interval censored time to first relapse. For the baseline
cocaine-use pattern, we consider both self-reported cocaine-use amount
trajectories and dichotomized use trajectories. Variations within the
generalized longitudinal trajectories are modeled through a latent
Gaussian process, which is characterized by a few leading functional
principal components. The association between the baseline longitudinal
trajectories and the time to first relapse is built upon the latent
principal component scores. The mean and the eigenfunctions of the
latent Gaussian process as well as the hazard function of time to first
relapse are modeled nonparametrically using penalized splines, and the
parameters in the joint model are estimated by a Monte Carlo EM
algorithm based on Metropolis--Hastings steps. An Akaike information
criterion (AIC) based on effective degrees of freedom is proposed to
choose the tuning parameters, and a modified empirical information is
proposed to estimate the variance--covariance matrix of the estimators.
\end{abstract}

%
\end{frontmatter}

\section{Introduction}\label{sec:introduction}

In cocaine dependence research, it has been shown that one's baseline
cocaine-use pattern is related to the risk of posttreatment cocaine relapse
[\citet{r10}],
along with many other factors such as cocaine withdrawal severity,
stress and negative mood
[\citet{r12}, \citeauthor{r28} (\citeyear{r28,r29})].
The timeline follow-back (TLFB) [\citet{r31}]
Substance Use Calendar is often used to retrospectively construct
trajectories of daily cocaine use in a baseline period before
treatment. The TLFB uses a calendar prompt and many other memory aids
(e.g., the use of key dates such as holidays, birthdays, newsworthy
events and other personal events as anchor points) to enhance the
accuracy of self-report substance-use estimates.
\citet{r8}
showed that the TLFB could provide reliable daily cocaine-use data that
had high retest reliability, high correlation with other cocaine-use
measures and high agreement with collateral informants' reports of
patients' cocaine use as well as results obtained from urine assays.

Based on the self-reported daily cocaine-use trajectories, certain
summary statistics can be derived and are often used as predictors in a
subsequent analysis to explain cocaine relapse outcomes. Commonly used
summary statistics include baseline cocaine-use frequency and average
daily use amount, and commonly used relapse outcome measures are time
to relapse (i.e., time to first cocaine use), frequency of use and
quantity of use per occasion during the follow-up period
[\citet{r4,r30}].
Among the different relapse outcome measures, time to first relapse
(which we also refer as ``relapse time'' for ease of exposition) is of
particular clinical importance because it signals the transition of a
cocaine-use pattern from abstinence to relapse.
\citet{r30}
examined time to cocaine relapse using Cox proportional hazards
regression models. They concluded that the amount of cocaine used per
occasion during the 90 days prior to inpatient admission was
significantly associated with relapse time. 
\citet{r11}
argued that because the baseline cocaine-use trajectories were random,
summary statistics derived from them were only estimates of one's
long-term cocaine-use behavior and could be subject to large
measurement error. In a regression setting, the use of error-prone
variables as predictors may cause severe bias to the regression coefficients
[\citet{r5}].
To mitigate the bias, \citet{r11} proposed a method-of-moments-based
calibration method for linear regression models and a subsampling
extrapolation method that is applicable to both linear and nonlinear
regression models. However, their methods require a restrictive
assumption that the baseline cocaine-use trajectories are stationary
processes, and their subsampling extrapolation method is an
approximation method which cannot completely eliminate the estimation
bias in survival analysis.

We propose a new modeling framework to link one's baseline cocaine-use
pattern to relapse time without assuming stationarity for the baseline
cocaine-use trajectories. We treat the baseline cocaine-use
trajectories as functional data [\citet{r23}]
and perform functional principal component analysis (FPCA) to these
trajectories. The resulting FPCA scores are then used as predictors to
model relapse time. We develop a joint modeling approach to conduct
FPCA and functional regression analysis simultaneously. We consider two
types of baseline cocaine-use trajectories: the first is the actual
self-reported daily cocaine-use amount as provided by the TLFB, whereas
the second is a dichotomized version of the first in the form of any
cocaine use versus no use. The actual daily cocaine-use amount can be
difficult to estimate depending on the length of the recalling period
and also due to the lack of a common scale to assess the amount used
for the different methods of consumption (e.g., intranasal use versus
injection). The dichotomized cocaine-use trajectories, although maybe
less informative, are subject to smaller errors and hence are more reliable.

There is a large volume of recent work on FPCA. See
\citet{r36,r13,r18}
for kernel-based FPCA approaches, and
\citet{r16}, Zhou, Huang and Carroll (\citeyear{r39,r40})
for spline-based FPCA methods. All these papers are concerned with the
Gaussian type of functional data and cannot be used for generalized
longitudinal trajectories.
\citet{r14}
proposed to model non-Gaussian longitudinal data by generalized linear
mixed models, where the FPCA can be performed with respect to some
latent random processes. Once the FPCA scores are obtained, a common
approach is to use them as predictors in a second-stage regression analysis
[e.g., \citet{r6,r37}].
As pointed out in
\citet{r19},
a potential problem with such an approach is that the estimation errors
in FPCA are not properly taken into account in the second stage
regression analysis, hence, the estimated coefficients can be biased
and variations in the estimators may be underestimated. By performing
FPCA and functional regression analysis simultaneously, we can avoid
these complications.

Our work is also related to joint modeling of longitudinal data and
survival time
[e.g., \citet{r24,r34,r123}, \citeauthor{r112} (\citeyear{r112,r35}),
\citet{r126}].
However, the vast majority of the existing literature focuses on the
instantaneous effect of longitudinal data on survival time. In other
words, the hazard rate of the event time is only related to the value
of the longitudinal process at the moment of event. In our problem, the
longitudinal trajectories were collected prior to the relapse period
and we want to use the entire baseline-use trajectory as a functional
predictor in the survival analysis. Survival analysis with functional
predictors is not well studied in the literature compared with other
functional regression models, and an extra complication in our data is
that the relapse time is interval censored (see Section~\ref{sec2.1} for
details). As noted in 
\citet{r3,r32},
one prominent difficulty in modeling interval censored survival data is
that, unlike right censored data, we cannot separate estimating the
baseline hazard function from estimating the hazard regression
coefficients using approaches such as the partial likelihood.
Therefore, we propose to model the log baseline hazard function as a
spline function. Some recent literature on spline models of the log
baseline hazard function for interval censored data includes
\citet{r3,r17,r25} and
\citet{r38}.


\section{Data structure and joint model}\label{sec2}

\subsection{Description of the motivating data}\label{sec2.1}

Our data came from a recently completed clinical trial for cocaine
dependence treatment. In the study, seventy-nine cocaine-dependent
subjects were admitted to the Clinical Neuroscience Research Unit
(CNRU) of the Connecticut Mental Health Center to receive an inpatient
relapse prevention treatment for cocaine dependence lasting for two to
four weeks. The CNRU is a locked inpatient treatment and research
facility that provides no access to alcohol or drugs and only limited
access to visitors.
Upon treatment entry, all subjects were interviewed by means of the
Structured Clinical Interview for DSM-IV
[\citet{r9}].
Variables collected during the interview include age, gender, race,
number of cocaine-use years and number of anxiety disorders present at
interview, among others. The TLFB Substance Use Calendar was used to
retrospectively construct daily cocaine-use history in the 90 days
prior to admission.

After completing the inpatient treatment, all participants were invited
back for follow-up interviews to assess cocaine-use outcomes. Four
interviews were conducted at days 14, 30, 90 and 180 after the
treatment. During each interview, daily cocaine-use records were
collected using the TLFB procedure for the period prior to the
interview date. A urine toxicology screen was also conducted to verify
the accuracy of a reported relapse or abstinence. A positive urine
sample test would suggest that the subject had used cocaine at least
once in the reporting period before the positive urine test, but the
test could not tell the exact cocaine-use date(s). If the self-reported
relapse time had no conflict with the urine tests, we consider it as an
observed event time. However, some subjects had reported no prior
cocaine use before the first positive urine sample test, their relapse
times were interval censored between their first positive urine test
and the previous negative test (if there was any). There were also
subjects who reported no cocaine use nor yielded any positive urine
samples for the entire study period. For these subjects, their relapse
time data were right censored at the last interview date.
In our data, about $50.6\%$ of the subjects had observed relapse time;
$31.6\%$ were interval censored and $17.8\%$ were right censored.

In what follows, let $N$ denote the number of study subjects. For the
$i$th subject, let $Y_i=\{Y_i(t_{ij}), j=1,\ldots, n_i\}$ be the
baseline cocaine-use trajectory, $T_i$ be a posttreatment relapse time
that may be right or interval censored, and $Z_i$ be an $m$-dimensional
covariate vector, where $t_{ij}$ is the $j$th observation time for the
$i$th subject within the baseline time interval $\CT$, $n_i$ is the
total number of such observation time, and $Z_i$ includes baseline
information on age, gender ($={}1$ for female and 0 for male), race
($={}1$ for African
American and 0 for the rest), number of cocaine-use years (Cocyrs) and
number of anxiety disorders present at the baseline interview
(Curanxs). As mentioned in the \hyperref[sec:introduction]{Introduction},
we consider
two cases that $Y_i(t)$ is either the self-reported use amount on day
$t$ or the dichotomized version.

\subsection{Modeling the baseline longitudinal trajectories}

\subsubsection{Generalized functional mixed model}
We assume that the longitudinal observations $Y_{ij}=Y_i(t_{ij})$ are
variables from the canonical exponential family 
[\citet{r22}]
with a probability density or mass function
%
\begin{equation}
\label{eq:exp_family} %
f(Y_{ij} | \theta_{ij}, \phi)=
\operatorname{exp} \biggl[ \frac{1}{a(\phi)}\bigl\{Y_{ij}
\theta_{ij}-b(\theta_{ij})\bigr\} +c(Y_{ij},\phi)
\biggr], %
\end{equation}
where $\theta_{ij}$ is the canonical parameter and $\phi$ is a
dispersion parameter. Denote $\mu_{ij}$ as the mean of $Y_{ij}$. Then
$\mu_{ij}$ is the first derivative of $b(\cdot)$ at $\theta_{ij}$, that
is, $\mu_{ij}=b^{(1)}(\theta_{ij})$. The inverse function of
$b^{(1)}(\cdot)$, denoted as $g(\cdot)$, is called the canonical link
function.
We consider two different types of trajectories: Gaussian trajectories
where $Y_i^{[1]}(t)= \log(0.5 +{}$ reported cocaine use on day $t$), and
dichotomized trajectories where $Y_i^{[2]}(t)=1$ if the $i$th subject
used cocaine on day $t$, and $=0$ otherwise. For Gaussian longitudinal
outcomes, $\theta_{ij}=\mu_{ij}$ and $f(Y_{ij} | \theta_{ij},\phi)$ is
the density of $\Normal(\theta_{ij},\phi)$; in the case of dichotomized
outcomes, $f(Y_{ij}| \theta_{ij}, \phi)$ is the binary probability mass
function with $\theta_{ij}=\operatorname{logit}\{P(Y_{ij}=1)\}$ and $\phi=1$.
We assume that $Y_i(t)$ is driven by a latent Gaussian process $X_i(t)$
such that $\theta_{ij}=X_i(t_{ij})$ and that $X_i(t)$ yields a standard
Karhunen--Lo\`eve expansion
%
\begin{equation}
\label{eq:KL_expansion} %
X_i(t)=\mu(t)+\psi(t)^{ T}
\xi_i  \qquad\mbox{for } t\in\CT, %
\end{equation}
where $\mu(t)=\E\{X_i(t)\}$ is the mean function, $\psi=(\psi
_1,\ldots,\psi_p)^{ T}$ is a vector of orthonormal functions also known
as the
eigenfunctions in FPCA, $\xi_i=(\xi_{i1},\ldots,\xi_{ip})^{ T}
\sim
\Normal(0, D_\xi)$ are the principal component scores, $D_\xi=\diag
(d_1,\ldots,d_p)$ and $d_1\ge d_2\ge\cdots\ge d_p >0$ are the
eigenvalues. In theory, the Karhunen--Lo\`eve expansion contains an
infinite number of terms, and truncating the expansion to a finite
order is a finite sample approximation to the reality. The number of
principal components $p$ becomes a model parameter and will be chosen
by a data-driven method.

\subsubsection{Reduced-rank model based on penalized B-splines}

We approximate the unknown functions $\mu(t)$ and $\psi(t)$ by B-splines
[\citet{r16,r39}].\vadjust{\goodbreak} The B-spline representation achieves two goals
simultaneously: smoothing and dimension reduction. Smoothing is needed
because the self-reported cocaine-use amount trajectories contain a
substantial amount of measurement error. With our spline
representation, each function is parameterized by a small amount of
spline coefficients and the estimates are further regularized by a
roughness penalty.

Let $\mathscr{B}(t)=\{\mathscr{B}_1(t),\ldots,\mathscr{B}_q(t)\}^T$ be a
$q$-dimensional B-spline basis defined on equally spaced knots in $\CT
$, $\theta_\mu$ be a
$q\times1$ vector and $\Theta_\psi=(\theta_{\psi1},\ldots, \theta
_{\psi p})$ be a $q\times p$ matrix of spline coefficients,
then the unknown functions are represented as
$\mu(t)=\mathscr{B}(t)^{ T}\theta_{\mu}$ and $\psi^{ T}
(t)=\mathscr
{B} (t)^{ T}\Theta_\psi$.
The general recommendation for choosing $q$ in the penalized spline
literature is to choose a relatively large number $q \gg p$, and let
the smoothness of the estimated functions be regularized by the
roughness penalty
[\citet{r26}].
The original B-spline basis functions are not orthonormal, therefore,
we employ the procedure prescribed by
\citet{r39}
to orthogonalize them so that $ \int\mathscr{B}(t)\mathscr
{B}(t)^T\,dt=I_q$, where $I_q$ is a $q\times q$ identity matrix. Under
this construction, the orthonormal constraints on $\psi(t)$ translate
into constraints on the coefficients, that is, $\Theta^T_\psi\Theta
_\psi=I_{p}$.
Then the reduced-rank model for the latent process takes the form
%
\begin{equation}
\label{eq:KL_expansion1} %
X_i(t)=\mathscr{B}(t)^T
\theta_{\mu}+\mathscr{B}(t)^T\Theta_\psi
\xi_i\qquad \mbox{subject to } \Theta^T_\psi
\Theta_\psi=I_p. 
%
\end{equation}

For the Gaussian trajectories, that is, the log-transformed cocaine-use
amount, $Y_i=B_i\theta_{\mu}+B_i\Theta_\psi\xi_i+\varepsilon_i$,
where $B_i=\{\mathscr{B}(t_{i1})^T, \ldots, \mathscr{B}(t_{in_i})^T\}
^T$ is the design matrix by interpolating the basis functions on the
observation time points and $\varepsilon_i \sim\Normal(0,\sigma
_\varepsilon^2 I_{n_i})$.
The conditional log-likelihood function for the baseline-use
trajectories is
%
\begin{eqnarray}
\label{eq:like_long1} %
\ell_{\mathrm{Long}}^{[1]}\bigl(
\Theta_{L}^{[1]}\bigr)= \sum
_{i=1}^{N} \ell _{\mathrm{Long}, i}^{[1]},
\nonumber
\\[-8pt]
\\[-8pt]
&&
\eqntext{\mbox{where } \displaystyle\ell_{\mathrm{Long}, i}^{[1]}=-
\frac{n_i}{2} \operatorname{log}\bigl(\sigma _{\varepsilon}^2\bigr)-
\frac{1}{2\sigma_{\varepsilon}^2} \|Y_i-B_i\theta_{\mu}-B_i
\Theta _\psi\xi_{i}\|^2,}
\end{eqnarray}
and $\Theta_{L}^{[1]}=(\theta_\mu^{ T}, \theta_{\psi
1}^{ T}
,\ldots,
\theta_{\psi p}^{ T},\sigma_\varepsilon^2)^{ T}$.

For the dichotomized trajectories, $\operatorname{log}\{\pi_{ij}/(1-\pi
_{ij})\}
=\SB^{ T}(t_{ij})\theta_{\mu}+\break  \SB^{ T}(t_{ij})\Theta
_\psi\xi_i$,
where $\pi_{ij}=P(Y_{ij}=1|\xi_i)$.
The conditional log-likelihood function is
%
\begin{eqnarray}
\label{eq:like_long2} %
\ell_{\mathrm{Long}}^{[2]}\bigl(
\Theta_{L}^{[2]}\bigr)= \sum
_{i=1}^{N} \ell _{\mathrm{Long}, i}^{[2]},
\nonumber
\\[-8pt]
\\[-8pt]
&& \eqntext{\mbox{where } \displaystyle\ell_{\mathrm{Long}, i}^{[2]}=\sum
_{j=1}^{n_i} \bigl\{ y_{ij}\operatorname{log}
\pi_{ij}+(1-y_{ij})\operatorname{log}(1-\pi_{ij}) \bigr\},}
\end{eqnarray}
and $\Theta_{L}^{[2]}=(\theta_\mu^{ T}, \theta_{\psi
1}^{ T}
,\ldots,
\theta_{\psi p}^{ T})^{ T}$.
To regularize the nonparametric estimators, we impose penalties on the
$L^2$ norms of their second derivatives 
[\citet{r7,r26}].
Define $\CJ_\SB=\int\mathscr{B}''(t)\* \mathscr{B}''(t)^T\,dt$, then
\[
\int \bigl\{\mu''(t) \bigr
\}^2 \,dt=\theta_\mu^{T}\CJ_\SB
\theta_\mu, \qquad \int \bigl\{\psi_k''(t)
\bigr\}^2 \,dt=\theta_{\psi l}^{
T}\CJ _\SB
\theta_{\psi l}. %
\]
%
The penalized log-likelihood for the baseline longitudinal data is
%
\begin{equation}
\label{eq:like_long_pen} %
\ell_{\mathrm{Long}}(\Theta_{L})-
\frac{1}{2} \Biggl( h_\mu\theta_\mu^{T}
\CJ_\SB\theta_\mu +h_\psi\sum
_{l=1}^{p} \theta_{\psi l}^T
\CJ_\SB\theta _{\psi l} \Biggr), %
\end{equation}
where $\ell_{\mathrm{Long}}$ is either $\ell_{\mathrm{Long}}^{[1]}$ or $\ell
_{\mathrm{Long}}^{[2]}$ for Gaussian and dichotomized trajectories,
respectively, and $h_\mu$ and $h_\psi$ are tuning parameters.

\subsection{Modeling the relapse time}

We assume that the relapse time $T_i$ depends on the baseline
cocaine-use history $Y_i(t)$ only through the latent process $X_i(t)$.
Moreover, the conditional hazard of $T_i$ given $\{X_i(t), t\in\CT\}
$ and the covariate vector $Z_i$ follows the Cox proportional hazards
model. Our way of including the functional covariate $X_i$ into
survival analysis is closely related to the functional linear model; see
\citet{r23,r37,r6,r19} and many others. More specifically, the
conditional hazard function of $T_i$ is
\[
\lambda_i(t| X_i,Z_i)=
\lambda_0(t) \exp\biggl\{ \int_\CT
X_i(s) \mathfrak{B}(s) \,ds+Z_i^T\eta\biggr
\}, %
\]
where $\lambda_0(t)$ is an unknown baseline hazard function, $\eta$ is
a coefficient vector and $\mathfrak{B}(s)$ is an unknown coefficient
function. 
When $X$ has the Karhunen--Lo\`eve expansion in (\ref{eq:KL_expansion}), the coefficient
function can be written as a linear combination of the eigenfunctions
$\mathfrak{B}(s)=\sum_{j=1}^p \beta_j \psi_j(s)$ and the integral in
the model can be simplified as $\int_\CT X_i(s) \mathfrak{B}(s)
\,ds=\sum_{j=1}^p \xi_{ij} \beta_j$, which motivates the model
%
\begin{equation}
\label{eq:cox} %
\lambda_i(t| \xi_i,Z_i)=
\lambda_0(t) \exp\bigl(\xi_i^T
\beta+Z_i^T\eta\bigr). %
\end{equation}
%

One important feature of the cocaine dependence treatment data is that
the relapse time is partially interval censored. That is, the data are
a mixture of noncensored, right censored and interval censored data.
For the subjects with interval censoring, we only know that the relapse
time occurred within an interval $[T_i^l,T_i^r]$, where $T_i^l\le
T_i^r$. We adopt the idea of
\citet{r3}
and model the log baseline hazard as a linear spline function
%
\begin{equation}
\label{eq:spline} %
\log\bigl\{\lambda_0(t)\bigr\}=
\mathbbm{a}_0+\mathbbm{a}_1 t +\sum
_{k=1}^K \mathbbm{b}_k (t-
\kappa_k)_+, 
%
\end{equation}
where $x_+\equiv\max(x,0)$ and $\kappa_k$'s are the knots. The spline
basis used in (\ref{eq:spline}) is also known as the truncated power basis
[\citet{r26}].
There are two immediate benefits for this model. First, $\lambda
_0(\cdot
)$ is guaranteed to be nonnegative, so that we do not have to consider
any constraints on the parameters when maximizing the likelihood.
Second, since $\log\lambda_0(\cdot)$ is modeled as a piecewise linear
polynomial, the cumulative hazard function $\Lambda_0(t)=\int_0^t
\lambda_0(u) \,du$ can be written out in an explicit form. For higher
order spline functions, such explicit expressions are not available.

To write out the likelihood for the relapse time, we use the following
notation. For the $i$th subject we observe $(T_i^l,T_i^r,\delta_i)$,
where $[T_i^l, T_i^r]$ gives the censoring interval and $\delta_i$ is
the indicator for right censoring. When $\delta_i=0$ and $T_i^l=T_i^r$,
the event time $T_i$ is right censored at $T_i^r$; when $\delta_i=1$
and $T_i^l<T_i^r$, $T_i$ is interval censored within $[T_i^l, T_i^r]$;
when $\delta_i=1$ and $T_i^l=T_i^r$, $T_i$ is observed at $T_i^r$. In
addition, $\delta_{0i}=I(\delta_i=1, T_i^l=T_i^r)$ is the indicator for
noncensored relapse time.
Denoting $\MX_i=(\xi_i^T,Z_i^T)^T$, the conditional log-likelihood
function for the relapse time is
[\citet{r3}]
%
%
\begin{eqnarray}
\label{eq:like_survive} %
\ell_{\mathrm{Relap}}(\Theta_S)&=&
\sum_{i=1}^N \ell_{\mathrm{Relap}, i}\qquad
\mbox{where}\nonumber
\\
\ell_{\mathrm{Relap}, i}&=& \delta_{0i} \bigl\{\operatorname{log}
\lambda_0\bigl(T_i^r\bigr)+\bigl(
\MX_i^T \theta \bigr) \bigr\} -(1-\delta_i)
\operatorname{exp}\bigl(\MX_i^T \theta\bigr)
\Lambda_0\bigl(T_i^r\bigr)
\\
&&{} +\delta_i(1-\delta_{0i})\operatorname{log} \bigl[
\operatorname{exp} \bigl\{\Lambda_0\bigl(T_i^r
\bigr)-\Lambda_0\bigl(T_i^l\bigr) \bigr\}
\operatorname{exp}\bigl(\MX_i^T \theta\bigr) \bigr],
\nonumber
\end{eqnarray}
and $\Theta_S=(\mathbbm{a}^{ T},\mathbbm{b}^{
T},\theta^{ T}
)^{ T}$
is the collection of parameters.

With the log baseline hazard function expressed as a linear spline
function, the log-likelihood function in (\ref{eq:like_survive}) can be
evaluated explicitly. To regularize the estimators, one commonly used
approach is to model the polynomial coefficients $\mathbbm
{a}=(\mathbbm
{a}_0,\mathbbm{a}_1)^T$ as fixed effects and the spline coefficients
$\mathbbm{b}=(\mathbbm{b}_1,\mathbbm{b}_2,\ldots,\mathbbm{b}_K)^T$ as
random effects with $\mathbbm{b}\sim\Normal(0, \sigma_\mathbbm{b}^2
I_{K})$. This mixed model setup leads to a penalized log-likelihood
%
\begin{equation}
\label{eq:like_survive pen} %
\ell_{\mathrm{Relap}}(\Theta_S)-
\frac{1}{2\sigma_\mathbbm{b}^2}{\mathbbm {b}}^T{\mathbbm{b}}. %
\end{equation}
%
\citet{r26} recommended to use a relatively large number of basis
functions in a penalized spline estimator, so that the smoothness of
$\log\lambda_0(\cdot)$ is mainly controlled by $\sigma_{\mathbbm{b}}^2$.
Following
\citet{r3},
we set $K=\min(\lfloor N/4 \rfloor,30)$, where $\lfloor x \rfloor$ is
the floor of $x$, and
choose the knots to be equally spaced with respect to the quantiles
defined on the unique values of $\{T_i^l,T_i^r,
(T_i^l+T_i^r)/2,i=1,\ldots,N\}$. The variance parameter $\sigma
_{\mathbbm{b}}^2$ is treated as a tuning parameter in our nonparametric
estimation. When analyzing the survival data alone,
\citet{r3}
proposed to select $\sigma_{\mathbbm{b}}^2$ by maximizing the marginal
likelihood using a Laplace approximation
[\citet{r2}].
Choosing $\sigma_{\mathbbm{b}}^2$ in our joint model is more
challenging and will be addressed in Section~\ref{sec:model_select}.

\subsection{The joint model}
The principal component scores $\xi_i$ of the longitudinal data are
also latent frailties in the survival model for the relapse time. By
imposing a normality assumption, the log-likelihood for $\xi$ is
%
\begin{equation}
\label{eq:like_frail} %
\ell_{\mathrm{Frail}}(\Theta_F)=\sum
_{i=1}^{N} \ell_{\mathrm{Frail}, i},\qquad
\ell_{\mathrm{Frail}, i}=- \tfrac{1}{2} \operatorname{log}|D_{\xi}|-
\tfrac{1}{2} \xi_i^T D_{\xi}^{-1}
\xi_i, %
\end{equation}
where $\Theta_F=(d_1,\ldots,d_p)^{ T}$ are the diagonal
elements of
$D_\xi$.

The complete data log-likelihood for the joint model is given by
combining the parts in (\ref{eq:like_long_pen}), (\ref
{eq:like_survive}) and (\ref{eq:like_frail}) as
%
\begin{equation}
\label{eq:joint_like} %
\ell_{C}(\Theta)=\sum
_{i=1}^N \ell_{C,i},\qquad
\ell_{C,i}= \ell_{\mathrm{Long},i}+\ell_{\mathrm{Relap},i}+\ell_{\mathrm{Frail},i},
\end{equation}
where $\Theta=(\Theta_{L}^{ T}, \Theta_{S}^{ T},
\Theta
_{F}^{ T}
)^{ T}$, and the penalized version of (\ref{eq:joint_like}) is
%
\begin{eqnarray}
\label{eq:joint_like_pen} %
&& \ell_P\bigl(\Theta; \xi, Y,
T^l,T^r,\delta,Z\bigr)
\nonumber
\\[-8pt]
\\[-8pt]
\nonumber
&& \qquad=\ell_{C}(\Theta)-\frac{1}{2\sigma_\mathbbm{b}^2}{\mathbbm {b}}^T{
\mathbbm {b}}-\frac{1}{2} \Biggl\{ h_\mu\theta_\mu^{T}
\CJ_\SB\theta_\mu +h_\psi\sum
_{l=1}^{p}\theta_{\psi l}^T
\CJ_\SB\theta_{\psi l} \Biggr\}.
\end{eqnarray}
Here $\xi$, $Y$, $T^l$, $T^r$, $\delta$ and $Z$ are the vectors or
matrices pooling the corresponding variables from all subjects.

\section{Methods}

\subsection{Model fitting by the MCEM algorithm}\label{sec:mcem}

We fit the joint model by an EM algorithm treating the latent variables
$\xi_i$ as missing values.
In our algorithm, we fix the tuning parameters $h_\mu$, $h_\psi$ and
$\sigma_\mathbbm{b}^2$ and focus on estimating the model parameters~$\Theta$. Selection of the tuning parameters is deferred to Section~\ref{sec:model_select}.

The loss function of the EM algorithm is
%
\begin{equation}
\label{eq:Q_EM} %
Q(\Theta; \Theta_{\mathrm{curr}})=\E \bigl\{
\ell_P\bigl(\Theta; \xi, Y, T^l,T^r,\delta,Z
\bigr) | Y, T^l,T^r,\delta,Z, \Theta_{\mathrm{curr}} \bigr
\}, %
\end{equation}
where $\ell_P$ is the penalized complete data log-likelihood in (\ref
{eq:joint_like_pen}) and $\Theta_{\mathrm{curr}}$ is the current value of
$\Theta
$. The algorithm updates the parameters by iteratively maximizing (\ref
{eq:Q_EM}) over $\Theta$. Given the complexity of the joint model, the
conditional expectation in (\ref{eq:Q_EM}) does not have a closed form,
we therefore approximate $Q(\Theta; \Theta_{\mathrm{curr}})$ by Markov Chain
Monte Carlo (MCMC). Let $\{\xi^{(1)},\ldots, \xi^{(R)}\}$ be MCMC
samples from the conditional distribution $(\xi_i | Y_i,
T^l_i,T^r_i,\delta_i,Z_i, \Theta_{\mathrm{curr}})$, and then $Q(\Theta;\Theta
_{\curr})$ can be approximated by $\wh Q(\Theta; \Theta_{\mathrm{curr}})
={1\over R} \sum_{k=1}^{R}\ell_P(\Theta; \xi^{(k)}, Y,
T^l,T^r,\delta,Z)$. This algorithm is a variant of the Monte Carlo EM
(MCEM) algorithm of \citet{r21}, and the details are provided in
Sections A.1 and A.2 of supplementary material [\citet{supp}].
To ensure convergence of the MCMC, we also monitor the Monte Carlo
error in the E-step using the batch means method of \citet{r106}. Specifically, we divide the Monte Carlo sequence $\{\xi^{(k)},
k=1,\ldots, R\}$ in to $R^{1/3}$ batches so that we have replicates of
$\wh Q(\Theta;\wh\Theta^{(s)})$ to evaluate the Monte Carlo error.

\subsection{Model selection by Akaike information criterion}\label
{sec:model_select}
The most pressing model selection issue in our joint model is to select
the number of principal components $p$ since it determines the
structure of the baseline trajectories and their association with the
relapse time. Another important issue is to select the tuning
parameters. As mentioned before, as long as we include enough of a
number of spline bases and place the knots reasonably, the performance
of the estimated functions is mainly controlled by the penalty
parameters $h_\mu, h_\psi$ and $\sigma_\mathbbm{b}^2$.
We propose to select $p$, $h_\mu, h_\psi$ and $\sigma_\mathbbm{b}^2$
simultaneously by minimizing an Akaike information criterion (AIC),
which is the negative log-likelihood plus a penalty on the model complexity.

In our setting, the log-likelihood on observed data requires
integrating out the latent variables $\xi$ from the complete data
likelihood (\ref{eq:joint_like}), which is intractable. A commonly used
approach is to replace the log-likelihood with its conditional
expectation given the observed data
[\citet{r15}].
Hence, the AIC is of the form
\[
\operatorname{AIC}\bigl(p, h_\mu, h_\psi,
\sigma_{\mathbbm{b}}^2\bigr)=
-2 \E
\bigl\{ \ell_C\bigl(\wh\Theta; \xi, Y, T^l,
T^r, \delta,Z\bigr) | Y, T^l, T^r, \delta,Z,
\wh\Theta \bigr\}+2 M, %
\]
where the conditional expectation is approximated by a Monte Carlo
average using the Monte Carlo samples in the last MCEM iteration and
$M$ is the effective degrees of freedom in the model.

For the longitudinal data, both the mean function $\mu(t)$ and the
eigenfunctions $\psi(t)$ are estimated by penalized splines. Following
\citet{r33},
the effective degrees of freedom for a P-spline estimator with a
penalty parameter $h$ is
\[
\mathrm{ df}(h)=\operatorname{ trace} \Biggl\{ \Biggl( \sum
_{i=1}^N  B_i^TB_i+h
\CJ_\SB \Biggr)^{-1} \sum
_{i=1}^N B_i^TB_i
\Biggr\}, %
\]
where $h$ can be either $h_\mu$ or $h_\psi$. Since our model consists
of one mean function and $p$ eigenvalues and eigenfunctions, the
effective degrees of freedom for the longitudinal data is $\mathrm
{df}(h_\mu)+ p \times\{\mathrm{df}(h_\psi)+1\}$.

Similarly, the effective degrees of freedom for the estimated log
baseline hazard function can be approximated by
[\citet{r26}]
\[
\mathrm{ df}\bigl(\sigma_\mathbbm{b}^2\bigr)=\operatorname{
trace} \Biggl\{ \Biggl( \sum_{i=1}^N {
\mathscr{T}_i^T\mathscr{T}_i}+
\frac{1}{\sigma
_\mathbbm{b}^2} \Biggr)^{-1} \sum_{i=1}^N
{\mathscr{T}_i^T\mathscr{T}_i} \Biggr\},
\]
where $\mathscr{T}_i$ is the design matrix from the truncated power
basis used in (\ref{eq:spline}). For interval censored subjects, we
approximate the event time by the midpoint $T_{i}^m$ of the interval
$[T_i^l,T_i^r]$ and the design matrix for the $i$th subject is
$
\mathscr{T}_i=\{(T_i^m-\kappa_1)_{+},\ldots, (T_i^m-\kappa_K)_{+}\}.
$
%

By taking into account the degrees of freedom in all model components,
the AIC for the joint model becomes
%
\begin{eqnarray}
\label{eq:aic} %
&&\operatorname{ AIC}\bigl(p,h_\mu,h_\psi,
\sigma_\mathbbm{b}^2\bigr)\nonumber
\\
&&\qquad =-2 \E \bigl\{ \ell_C\bigl(\wh\Theta; \xi, Y, T^l,
T^r, \delta,Z\bigr) | Y, T^l, T^r, \delta,Z,
\wh\Theta \bigr\}
\\
&&\qquad\quad{} +2 \bigl[\mathrm{ df}(h_{\mu})+p\times\bigl\{\mathrm{ df}(h_\psi)
+1
\bigr\}+\mathrm{ df}\bigl(\sigma _{\mathbbm{b}}^2\bigr)+m+p \bigr].
\nonumber
\end{eqnarray}
Searching for the minimum of AIC in a four-dimensional space is
extremely time consuming. One possible simplification is to assume that
the baseline mean and eigenfunctions have about the same roughness and
set $h_\mu=h_\psi\equiv h $. Then for each value of $p$, we search for
the optimal value of $h$ and $\sigma_{\mathbbm{b}}^2$ over five grid
points in each dimension. We adopt this search scheme in all of our
numerical studies and it proves to be computationally feasible.



%

\subsection{Variance estimation}\label{sec:inference}

To make inference on parameters in the joint model, we need to estimate
the variance--covariance matrix of the estimator $\wh\Theta$. Let $\CO
=(Y, T^l, T^r, \delta, Z)$ be the observed data.
\citet{r20}
showed that the covariance matrix of $\wh\Theta$ can be approximated
by the inverse of the observed information matrix
%
\begin{eqnarray}
\label{eq:louis} %
I_\Theta&=&- \E \biggl\{ {\partial^2 \over\partial\Theta\,\partial
\Theta
^{ T}}
\ell_P(\Theta;\xi,\CO)\Big | \CO \biggr\}\nonumber
\\
&&{} - \E \biggl\{{\partial\over\partial\Theta} \ell_P(\Theta;\xi,\CO)
{\partial\over\partial\Theta^{ T}} \ell_P(\Theta;\xi,\CO) \Big| \CO \biggr\}
\\
&&{} + \E \biggl\{{\partial\over\partial\Theta} \ell_P(\Theta;\xi,\CO ) \Big|
\CO \biggr\} \E \biggl\{{\partial\over\partial\Theta
^{ T}} \ell_P(\Theta;\xi,
\CO) \Big|\CO \biggr\} ,
\nonumber
\end{eqnarray}
where $\ell_P$ is the penalized log-likelihood based on complete data
(\ref{eq:joint_like_pen}).
We can estimate this information matrix by evaluating the partial
derivatives at the final estimator $\wh\Theta$ and replacing the
conditional expectations by Monte Carlo averages using the Monte Carlo
samples generated in the final EM iteration. 

One important distinction between our model and the generalized linear
mixed models or other joint models is that the eigenfunctions are not
identifiable without the orthonormal constraints in (\ref{eq:KL_expansion1}).
Because of the constraints, the real number of free parameters in
$\Theta_{\psi}$ is lower than the nominal dimension. As a result, the
information matrix defined above might be singular. One solution is to
reparameterize $\Theta_{\psi}$ so as to remove the constraints. Details
are given in supplementary material [\citet{supp}].

A referee pointed out the methods by \citet{r118} and \citet{r119} can
also be used to estimate the asymptotic variance of $\wh\Theta$. These
methods are not only based on observed information, but also evaluate
the derivatives numerically by running additional Markov chains. It is
worth pointing out that these methods are designed for the cases where
there is no constraint on the parameter~$\Theta$. Extending these
methods to our problem calls for future research.


%

\section{Simulation study}\label{sec:simulation}
We illustrate the performance of the proposed methods by a simulation
study. To mimic the real data, we consider two simulation settings
where the baseline longitudinal trajectories are Gaussian and binary,
respectively. In both settings, we simulate $N=100$ independent
subjects, with $n_i=20$ baseline longitudinal observations equally
spaced on the time interval $\CT=[0,20]$.

Gaussian baseline trajectories are generated as
$Y_i(t)=X_i(t)+\varepsilon_i(t)$, where $X_i(t)$ is the $i$th
realization of a Gaussian process with the Karhunen--Lo\`eve expansion
(\ref{eq:KL_expansion}). We let the mean function be $\mu
(t)=t/60+\operatorname{sin}(3\pi t/20)$, the eigenvalues be $d_1=9$,
$d_2=2.25$ and $d_k=0$ for $k\geq3$, and the eigenfunctions be $\psi
_1(t)=-{\cos}(\pi t/10)/\sqrt{10}$, $\psi_2(t)={\sin}( \pi
t/10)/\sqrt
{10}$. The principal component scores are simulated as $\xi_i=(\xi
_{i1},\xi_{i2})^{ T}\sim\Normal({ 0}, D_\xi)$ with $D_\xi=
\rm
{diag}(9,2.25)$.
The error $\varepsilon(t)$ is a Gaussian white noise process with
variance $\sigma_{\varepsilon}^2=0.49$. In the case of the binary
baseline, $Y_{ij}$ are generated from a Bernoulli distribution with the
probability $g^{-1}\{X_i(t_{ij})\}$, where the latent process $X$ is
simulated the same way as for the Gaussian baseline trajectories and
$g(\pi)=\operatorname{ log}(\frac{\pi}{1-\pi})$ for $0<\pi<1$.

\begin{figure}[b]

\includegraphics{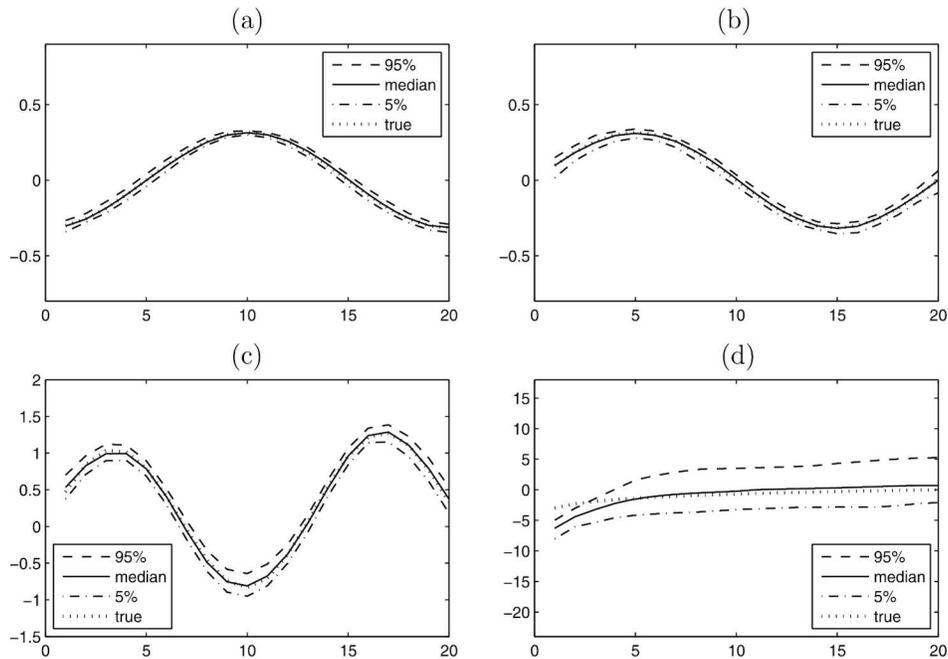}

\caption{Summary of the nonparametric estimators in the simulation
study when the baseline longitudinal trajectories are Gaussian. The
four panels correspond to $\wh\psi_1(t)$, $\wh\psi_2(t)$, $\wh\mu(t)$
and the log baseline hazard function, respectively. In each panel, the
dotted curve is the true function, the solid curve is the median of the
estimator, the dash-dot and dashed curves are the 5\% and 95\%
pointwise percentiles. \textup{(a)} 1st eigenfunction. \textup{(b)} 2nd eigenfunction.
\textup{(c)} Baseline mean function. \textup{(d)} Log baseline hazard function.}
\label{fig:simu_gauss}
\end{figure}

Under both simulation settings, we simulate the failure time $T_i$ from
the Cox proportional hazards model (\ref{eq:cox}), which includes the
effects of the principal component scores and a covariate $Z_i$. We let
$Z_i$ be a binary random variable with a success probability of 0.5,
the regression coefficients be $\theta=(\beta^T, \eta)^T=(1,1,1)^T$,
and the baseline hazard function be $\lambda_0(t)=t/20$ for $t\geq0$.
We assume that the failure time is interval censored at random and set
the censoring time to be $4$, $10$ and $20$.
Let the censoring indicator $\delta_i$ be a binary variable independent
of $\xi_i$ and $Z_i$ with $P(\delta_i=1)=0.5$. When $\delta_i=1$, the
event time $T_i$ is censored in the interval between the two closest
censoring time; if $T_i$ is less than 4, it is censored in
$[T_i^l=0, T_i^r=4]$; if $T_i$ is over 20, it is automatically right
censored at 20. Overall,
the data structure is similar to the cocaine dependence treatment data
described in Section~\ref{sec2}:
about $12\%$ of the failure times are right censored, $43\%$ are
interval censored, and the remaining $45\%$ are observed.

\begin{figure}

\includegraphics{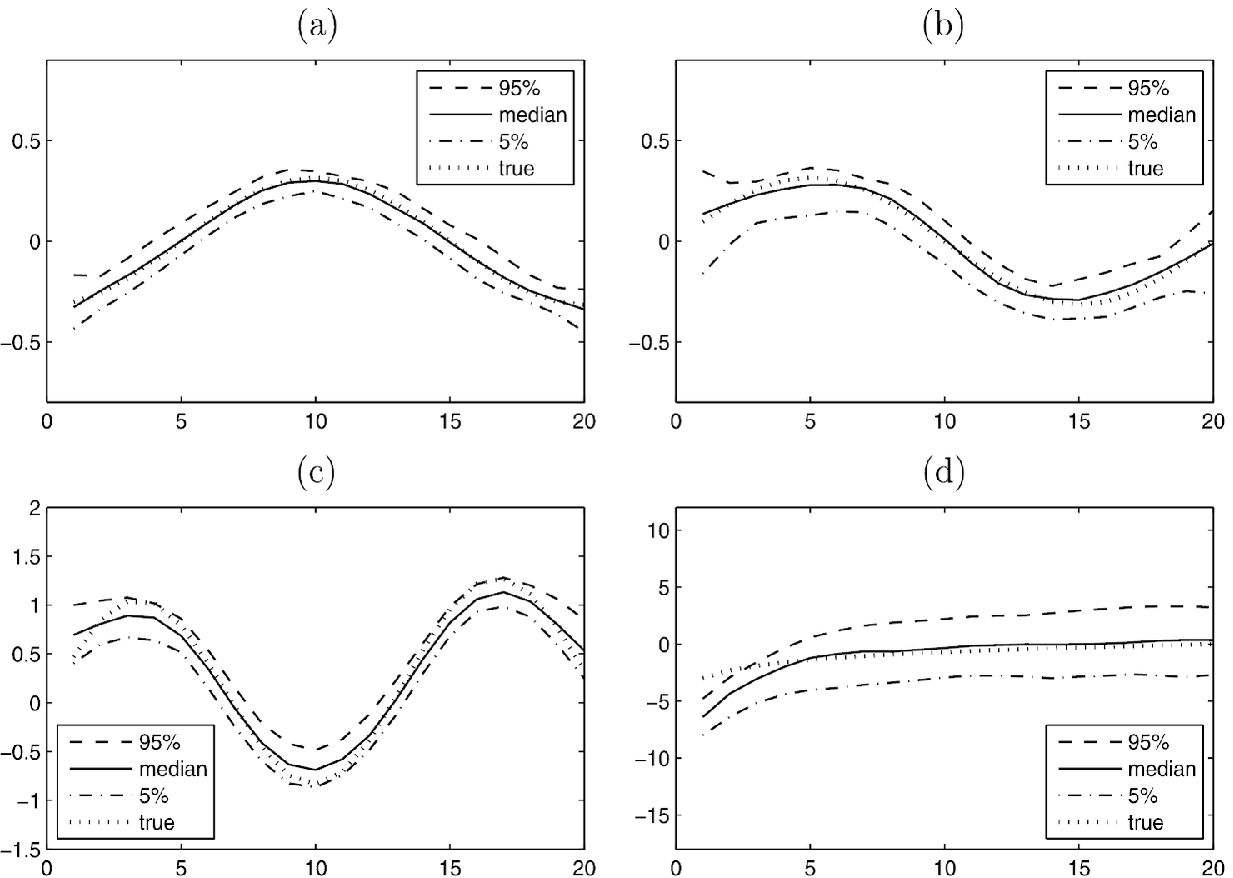}

\caption{Summary of the nonparametric estimators in the simulation
study when the baseline longitudinal trajectories are binary. The four
panels correspond to $\wh\psi_1(t)$, $\wh\psi_2(t)$, $\wh\mu(t)$ and
the log baseline hazard function, respectively. In each panel, the
dotted curve is the true function, the solid curve is the median of the
estimator, the dash-dot and dashed curves are the 5\% and 95\%
pointwise percentiles.
\textup{(a)} 1st eigenfunction. \textup{(b)} 2nd eigenfunction.
\textup{(c)} Baseline mean function. \textup{(d)} Log baseline hazard function.}
\label{fig:simu_binary}
\end{figure}

For both baseline settings, we repeat the simulation 100 times and
apply the proposed method to fit the joint model.
For the results reported below, we use $q=8$ cubic B-splines to model
the mean and eigenfunctions of the latent longitudinal process and
$K=12$ spline basis functions to model the log baseline hazard
function. Our experience and those of many others
[e.g., \citet{r3,r26,r39}]
suggest that the performance of penalized spline estimators is mainly
controlled by the penalty parameters and is not sensitive to the choice
of spline basis.

To choose the number of principal components $p$ and the penalty
parameters $h_\mu$, $h_\psi$ and $\sigma_\mathbbm{b}^2$, we conduct a
grid search using the proposed AIC (\ref{eq:aic}). For all the
simulations, the AIC selects the correct number $p=2$ of principal
components about 77\% of the time and selects $p=3$ for the remaining
23\% of the time. Since AIC has a well-known tendency to select an
over-fitted model and over-fitting is in general considered less
problematic than under-fitting, this performance is quite satisfactory.
For the estimation results below, we use the penalty parameters
selected by AIC when $p$ is fixed at 2.\looseness=-1

We summarize in Figures~\ref{fig:simu_gauss} and \ref{fig:simu_binary} the nonparametric estimators when the
baseline longitudinal trajectories are Gaussian and binary,
respectively. Each figure contains four panels that summarize $\wh\psi
_1(t)$, $\wh\psi_2(t)$, $\wh\mu(t)$ and the log baseline hazard
function. We show in each panel the true curve, the median, and the
$5$th and 95th pointwise percentiles of the
estimators. As we can see, the spline estimators perform very well in
both simulation settings, and the median and the pointwise percentiles
of the estimated curves are very close to the truth. Between the two
types of baseline longitudinal data, binary trajectories are less
informative, and hence the estimated curves are more variable. For
instance, the integrated mean squared error for the two eigenfunctions
are 0.0072 and 0.0150 in the Gaussian case and are 0.0462 and 0.1206 in
the binary case. The true log hazard function is $\log(t/20)$, which is
$-\infty$ at $t=0$; this explains the bigger bias of our spline
estimator near 0. The bias in the nonparametric part has little effect
on estimation of the parametric components such as $\theta$.

%
\begin{table}
\caption{Estimation results of the parametric components under both simulation
settings, with either Gaussian or binary baseline trajectories. 
Presented in the table are the true value of the parameters, mean and
Monte-Carlo standard deviations (Stdev) of the estimated parameters,
and the mean of the estimated standard error using the Louis formula
(Stder). The joint modeling method (joint) is the proposed method, and
the two-stage method is by plugging estimated FPCA scores into a second
stage survival analysis}\label{table:nonlin} 
\begin{tabular*}{\textwidth}{@{\extracolsep{\fill}}lccccccc@{}}
\hline
\textbf{Method} &\textbf{Parameter} & $\bolds{\beta_1}$ &
$\bolds{\beta_2}$ &$\bolds{\eta}$ &$\bolds{d_1}$ & $\bolds{d_2}$ &
$\bolds{\sigma_\varepsilon^2}$ \\
\hline
\multicolumn{4}{@{}l}{Gaussian baseline trajectory} \\
Two-stage&True & 1.0000 & 1.0000 &1.0000& 9.0000 &2.2500 &0.4900 \\ 
 &Mean & 0.8154 & 0.8092 &0.7972& 8.9248 &2.0224 &0.4443 \\
&Stdev & 0.0911 & 0.1513 &0.3302& 1.1193 &0.3183 &0.0147 \\
Joint &Mean & 0.9824 & 1.0130 &0.9782& 9.1184 &2.0861 &0.4839 \\
&Stdev & 0.1253 & 0.1926 &0.3885& 1.1558 &0.3349 &0.0157 \\
&Stder & 0.1184 & 0.1593 &0.3469& 1.3661 &0.3633 &0.0154 \\[6pt]
\multicolumn{4}{@{}l}{Binary baseline trajectory} \\
Two-stage&True & 1.0000 & 1.0000 &1.0000& 9.0000 &2.2500& \\ 
 &Mean & 0.8187 & 0.6681 &0.4642& 6.4365 &2.1158& \\
&Stdev & 0.1759 & 0.4840 &0.2658& 1.1685 &0.5384& \\ 
Joint &Mean & 0.9798 & 0.9890 &0.9997& 9.3307 &2.2823& \\
&Stdev & 0.1380 & 0.1727 &0.3724& 1.9894 &0.8342& \\
&Stder & 0.1192 & 0.1553 &0.3412& 2.0035 &0.6059& \\
\hline
\end{tabular*}
\end{table}

We summarize the estimation results of the parametric components for
both settings in Table~\ref{table:nonlin}, where we show the means and Monte Carlo
standard deviations of the estimators. As we can see, the estimators
for the parametric components are approximately unbiased and the
standard deviations are reasonably small. We also present the means of
the estimated standard errors using the modified empirical information
in Section~\ref{sec:inference}, and find that the standard errors
slightly underestimate the true standard deviations. This
underestimation of standard error is quite common in semiparametric
models under small sample sizes, since the standard error is based on
an estimate of the asymptotic variance, which only captures the leading
term in the asymptotic distribution of the point estimator
[\citet{r125}].


To demonstrate the advantage of the joint modeling approach, we also
provide a comparison between our method and a two-stage functional
survival analysis approach, where we perform FPCA to the longitudinal
trajectory first and then use the estimated principal component scores
as predictors in the second-stage survival analysis. For Gaussian
longitudinal trajectories, the FPC scores are estimated by the
principal analysis by the conditional expectation (PACE) method [\citet
{r36}]; for the dichotomized trajectories, the FPC scores are estimated
by the method of \citet{r14} which is implemented in a PACE-GRM package
in Matlab. The estimation results of the two-stage estimator are also
provided in Table~\ref{table:nonlin}. We can see that the two-stage estimators for
$\beta
$ and $\eta$ are severely biased. This bias is the result of the
attenuation effect caused by the estimation errors in the FPC scores.

\section{Cocaine dependence treatment data}\label{sec:data}

We apply our proposed joint modeling approach to analyze the cocaine
dependence treatment data described in Section~\ref{sec2}. For the baseline
cocaine-use trajectories, we consider both the (log-transformed)
cocaine-use amount trajectories and the dicho\-tomized trajectories.
Relapse time is determined from the self-reported posttreatment
cocaine-use trajectories as well as the urine sample tests. As we
discussed in Section~\ref{sec2}, the relapse time is partially interval/right
censored. We use the five covariates described in Section~\ref{sec2} in the Cox
model, that is, age, gender, race, Cocyrs and Curanxs. To capture
potential weekly periodic patterns of the baseline trajectories, we
aligned the baseline trajectories by weekdays such that all
trajectories start from the first Sunday of the baseline period and
last for 80 days.

We use 30 cubic B-spline basis functions to model the mean and
eigenfunctions of the baseline trajectories so that there are about two
knots within each weak and the basis functions are flexible enough to
capture possible weekly patterns in the data. The smoothness of these
nonparametric estimators are governed by the data-driven tuning
parameters. We use 12 linear spline basis functions to model the
baseline hazard function, similar to the choice in \citet{r11}.
We choose the number of principal components and the penalty parameters
$h_\mu, h_\psi$ and $\sigma_\mathbbm{b}^2$ by the proposed AIC. The AIC
selects three principal components for both types of baseline
trajectories. The estimated eigenvalues are 16.1960, 2.2097 and 0.8673
for the cocaine-use amount trajectories and 61.3838, 0.8986 and 0.1695
for the dichotomized trajectories.

\begin{figure}

\includegraphics{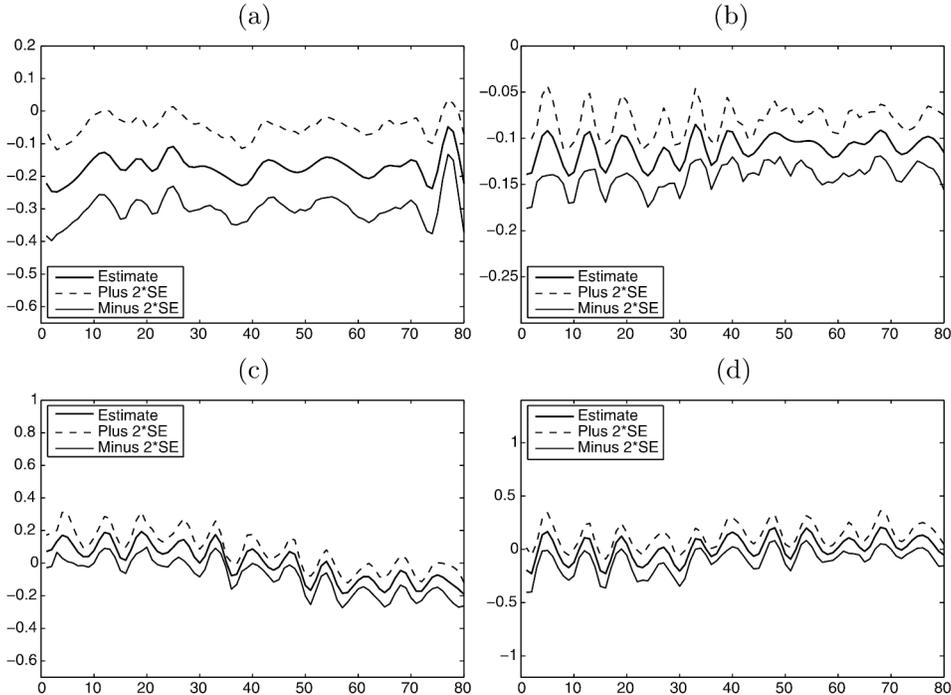}

\caption{The mean function and the first three eigenfunctions for the
cocaine-use amount trajectories.
\textup{(a)} Mean function. \textup{(b)} 1st eigenfunction.
\textup{(c)} 2nd eigenfunction.
\textup{(d)} 3rd eigenfunction.}
\label{fig:pc_effect_data}
\end{figure}

\begin{figure}

\includegraphics{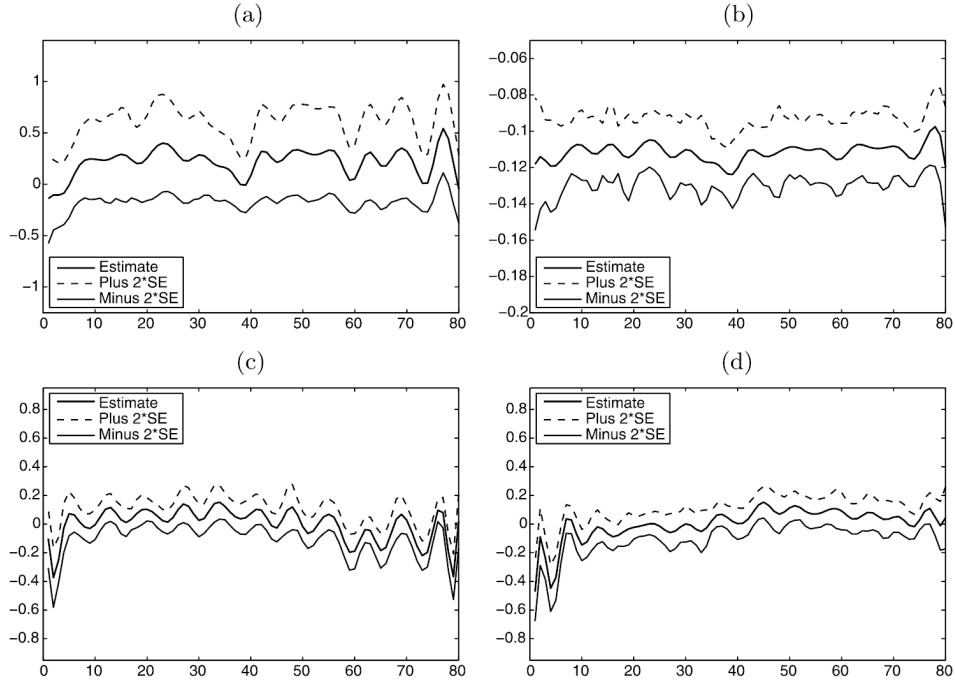}

\caption{The mean function and the first three eigenfunctions for the
latent process of the dichotomized trajectories.
\textup{(a)} Mean function. \textup{(b)} 1st eigenfunction.
\textup{(c)} 2nd eigenfunction.
\textup{(d)} 3rd eigenfunction.}\label{fig:pc_effect_data2}
\end{figure}

We show the estimated mean and eigenfunctions for the cocaine-use\break
amount trajectories in Figure~\ref{fig:pc_effect_data} and for the
dichotomized trajectories in Figure~\ref{fig:pc_effect_data2}. The
curves estimated from the two types of trajectories exhibit rather
similar patterns, and they all show clear weekly periodic structures---the
baseline trajectories contain 11 weeks of data and these curves
have 11 peaks and troughs matching the weekdays rather closely. If we
look beyond the local periodic structures and focus on the overall
trend of these curves over the entire baseline period, we can see that
the mean functions are reasonably flat except near the beginning and
the end of the baseline period. The overall trend in the first
eigenfunction is a negative constant function. Increasing the loading
on the first principal component leads to less cocaine use (or lower
use probability for dichotomized trajectories), and hence the score on
the first principal component represents the overall use amount (or
probability) of a patient.
The second principal component represents an overall decreasing trend
in use amount (or probability) over the recall period. The third
principal component is a higher order nonlinear trend in the trajectories.

To confirm that the weekly structures in these curves are real, we also
provide pointwise standard error bands in the plots. Since our
simulation study shows that the standard error based on the Louis
formula underestimates the true standard deviation under a small sample
size, we estimate the standard error using a bootstrap procedure
instead. In our bootstrap procedure, we resample the subjects with
replacement, fit the joint model to the bootstrap samples using the
same tuning parameters as for the real data, and estimate the standard
deviations of the estimators using their bootstrap replicates
pointwisely. The confidence bands in Figures~\ref{fig:pc_effect_data}
and \ref{fig:pc_effect_data2} are based on 100 bootstrap replicates.
These confidence bands confirm that the weekly structures in the
eigenfunctions are real. Note that the confidence bands in Figure~\ref{fig:pc_effect_data2} are wider than those in Figure~\ref{fig:pc_effect_data} because the dichotomized trajectories are less informative.


The estimated regression coefficients for the Cox model and the
corresponding standard errors and $p$-values are reported in Table~\ref{tab:data}. The standard errors are obtained by bootstrap with 100
replicates. For both types of baseline trajectories, the second
principal component has a significant positive effect on the hazard
rate of relapse time. This suggests that patients with a decline in
recent cocaine-use amount or
probability relapsed faster. Subjects who experienced such a decline might
have established a longer period of abstinence before entering treatment
than those who did not. As a result, it would not be surprising for the onset
of their cocaine withdrawal symptoms to start sooner; this could have in
turn caused a faster relapse. Among the covariates, Cocyrs is
significant, suggesting subjects who had used cocaine for fewer years
tended to relapse later.

For comparison purposes, we also report in Table~\ref{tab:data} the
estimation result of the two-stage procedure described in Section~\ref{sec:simulation}. In this procedure, FPCA and survival analysis are
done in successive steps, and the estimation errors in the estimated
principal component scores are not properly taken into account in the
survival analysis. It is not surprising that the estimation
coefficients for the principal component scores by the two-stage
procedure are attenuated and none of them are significant.

\begin{table}
\tabcolsep=0pt
\caption{Cocaine data analysis under the joint model using either the
cocaine-use amount trajectories (Amnt.) or the dichotomized use
trajectories (Dich.).
The table shows the estimated coefficients for the variable $\xi$ and
five covariates.
Cocyrs and Curanxs denote the number of cocaine-use years and the
number of current anxiety symptoms at baseline interview, respectively.
``Stder'' is the estimated standard error, which is calculated under
bootstrap in the joint model.
The $p$-value with $^*$ indicates significance at $\alpha=0.05$ level} \label{tab:data}
\begin{tabular*}{\textwidth}{@{\extracolsep{\fill}}lcd{1.5}d{2.4}d{2.4}d{2.4}d{2.4}d{1.5}c@{}}
\hline
\textbf{Amnt.} &
\multicolumn{1}{c}{$\bolds{\xi_1}$} & \multicolumn{1}{c}{$\bolds{\xi_2}$} &
\multicolumn{1}{c}{$\bolds{\xi_3}$} &
\multicolumn{1}{c}{\textbf{Gender}} & \multicolumn{1}{c}{\textbf{Race}} &
\multicolumn{1}{c}{\textbf{Age}} & \multicolumn{1}{c}{\textbf{Cocyrs}} &
\multicolumn{1}{c@{}}{\textbf{Curanxs}} \\
\hline
\multicolumn{3}{@{}l}{Two-stage estimator} & \\
Est &0.0418 &0.1616 &-0.2251 &-0.3818 & -0.4081 &-0.0467 &0.1182 &
0.2664 \\
Stder &0.0316 &0.0995 &0.1590 & 0.2986 & 0.3305 &0.0276 &0.0347 &
0.2584 \\
$p$-value &0.1870 &0.1046 &0.1570 & 0.2011 & 0.2169 &0.0908
&0.0007^* & 0.3024\\[3pt]
\multicolumn{3}{@{}l}{Joint model} & \\
Est &0.0420 &0.1802 &-0.2021 &-0.3255 &-0.3343 &-0.0449 &0.1098 &0.2348
\\
Stder &0.0352 &0.0867 &0.2394 & 0.3462 & 0.2591 &0.0342 &0.0407
&0.2109 \\
$p$-value &0.2327 &0.0377^{*} &0.3985 & 0.3471 & 0.1969 &0.1895
&0.0070^{*} &0.2655 \\
[0.5ex] 
\hline
\textbf{Dich.} &
\multicolumn{1}{c}{$\bolds{\xi_1}$} & \multicolumn{1}{c}{$\bolds{\xi_2}$} &
\multicolumn{1}{c}{$\bolds{\xi_3}$} &
\multicolumn{1}{c}{\textbf{Gender}} & \multicolumn{1}{c}{\textbf{Race}} &
\multicolumn{1}{c}{\textbf{Age}} & \multicolumn{1}{c}{\textbf{Cocyrs}} &
\multicolumn{1}{c@{}}{\textbf{Curanxs}} \\
\hline
\multicolumn{3}{@{}l}{Two-stage estimator} &\\
Est &0.0008 &0.0131 &-0.1331 &-0.3538 & -0.2664 &-0.0437 &0.1031 &
0.3582 \\
Stder &0.0137 &0.0762 &0.1158 & 0.2743 & 0.2919 &0.0223 &0.0306 &
0.2802 \\
$p$-value &0.9552 &0.8636 &0.2501 &0.8030 & 0.3613 &0.0500
&0.0007^{*} & 0.2011\\[3pt]
\multicolumn{3}{@{}l}{Joint model} &\\
Est &0.0064 &0.1840 &-0.2344 &-0.3536 &-0.1567 &-0.0408 &0.0947 &0.2431
\\
Stder &0.0135 &0.0936 &0.2261 & 0.3128 &0.2343 &0.0315 &0.0393 &0.2544
\\
$p$-value &0.6339 &0.0493^{*} &0.3000 & 0.2583 &0.5035 &0.1951
&0.0160^{*} &0.3392 \\
\hline
\end{tabular*}
\end{table}

Following a referee's suggestion, we have also performed PCA to the use
amount trajectories without B-spline representation and roughness
penalty regularization and use the PC scores in the survival analysis.
The estimated Cox regression coefficients for the first three principal
components are $(0.0380,0.0218,-0.0169)$ with standard errors
$(0.0562,0.1283,0.1738)$. In other words, none of these PC scores is
found to be significantly related to the first relapse time. This is
because the cocaine-use amount trajectories contain a large amount of
error (due to self-reporting and converting different consumption
methods to equivalent grams), and without regularization and joint
modeling the estimation errors in the PC scores greatly attenuate the
Cox regression coefficients and reduce statistical power. Such a direct
PCA approach is not applicable to the dichotomized trajectories.

In our joint modeling analysis, we also closely monitor the convergence
of the Markov Chain. We estimate the Monte Carlo error in the final EM
iteration using the method described in Section~\ref{sec:mcem}, which
is $8.3408 \times10^{-4}$ for the cocaine-use amount trajectories and
$7.8830 \times10^{-4}$ for the dichotomized trajectories.

In a previous work, \citet{r30}
analyzed a similar data set and concluded that the baseline average
cocaine-use amount had a significant negative effect on the hazard
function of relapse; this implies that those who used less during the
baseline period tended to relapse sooner, which is counterintuitive.
In \citet{r11}, the authors
argued that the counterintuitive results could be due to measurement
error in the average use amount. After having accounted for the
measurement error, they found that the baseline average cocaine-use
amount was no longer significant. Since the first principal component
in our joint model is closely related to the baseline average
cocaine-use amount, our result further confirms the analysis of
\citet{r11}.
However, we have also found that the subject-specific decreasing trend
in the cocaine-use trajectories (i.e., the second principal component)
is 
related to faster relapse, while such a finding was not made by either
\citet{r30} or \citet{r11}.

\section{Summary}

In studying the relationship between baseline cocaine-use patterns and
posttreatment time to first cocaine relapse, most existing literature
only makes use of some basic summary statistics derived from the
cocaine-use trajectories, such as the average use amount and frequency
of use. These summary statistics are subject to measurement error and
cannot fully describe the dynamic structure of the baseline trajectories.

We propose an innovative joint modeling approach based on functional
data analysis to jointly model the baseline generalized longitudinal
trajectories and the interval censored failure time. Specifically, we
model the latent process that drives the longitudinal responses as
functional data, approximate the mean and eigenfunctions of the latent
process by flexible spline basis functions, and propose a data-driven
method to determine the number of principal components and hence the
covariance structure of the longitudinal data. We propose and implement
a Monte Carlo EM algorithm to fit the model and modified empirical
information to estimate the standard error of the regression
coefficients. Our analysis of the cocaine dependence treatment data
shows that the relapse time is related to a decreasing trend in the
cocaine-use behaviors rather than the average use amount.

Our proposed model can also be used to predict the first relapse time
of the new subject. For a future subject, suppose that we only observe
his/her baseline cocaine-use amount trajectory $\{Y^*(t), t\in\CT\}$,
then we can predict his/her first relapse time $T^*$ using an empirical
Bayes method. Using the proposed joint model, we can write out the
conditional distribution $[T^*, \xi^* | Y^*(t), t\in\CT]$, where
$\xi
^*$ is the vector of latent principal component scores for the new
subject. We can use the model parameters estimated from the training
data set, and run an MCMC to draw samples from this conditional
distribution. We use the MCMC samples to estimate the posterior
distribution of $T^*$, which provides both a point predictor and
prediction intervals.

As all Monte Carlo based methods, our methods are computationally
intense. For the cocaine dependence treatment data, it takes about 25
EM iterations for the algorithm to converge and the running time is
about 1.5 hours using the self-reported use amount trajectories and
about 2.5 hours using the dichotomized use trajectories. It takes a lot
longer to perform model selection and bootstrap, since we have to fit
the model many times. However, we argue that the computation time is a
worthy price to pay in exchange for unbiased estimates and correct
statistical inference. One of our future research directions is to
accelerate the EM algorithm using graphics processing units (GPU) and
parallel computing.

\section*{Acknowledgments}
We thank the Editor, the Associate Editor and three anonymous referees
of an earlier version of this paper who gave valuable advice on
clarifying and explaining our ideas.

\begin{supplement}[id=suppA]
\stitle{Supplement A}
\slink[doi]{10.1214/15-AOAS852SUPP} 
\sdatatype{.pdf}
\sfilename{aoas852\_supp.pdf}
\sdescription{The online supplementary material for this paper contains
the technical details of the MCEM algorithm to fit the model,
estimation of the covariance matrix of the estimator, additional
simulation results and sensitivity analysis in the real data analysis.}
\end{supplement}

%

\printaddresses
\end{document}